\documentclass[11pt,preprint]{aastex}
\usepackage{apjfonts}

\def\gax{\mathrel{\raise.3ex\hbox{$>$}\mkern-14mu\lower0.6ex\hbox{$\sim$}}}
\def\lax{\mathrel{\raise.3ex\hbox{$<$}\mkern-14mu\lower0.6ex\hbox{$\sim$}}}
\def\gtorder{\mathrel{\raise.3ex\hbox{$>$}\mkern-14mu
             \lower0.6ex\hbox{$\sim$}}}
\def\ltorder{\mathrel{\raise.3ex\hbox{$<$}\mkern-14mu
             \lower0.6ex\hbox{$\sim$}}}

\begin{document}

\title{On Absorption by Circumstellar Dust,  \\ 
    With the Progenitor of SN~2012aw as a Case Study}

\author{
   C.~S. Kochanek$^{1,2}$, R. Khan$^{1,2}$ \& X. Dai$^{3}$ 
  }

\altaffiltext{1}{Department of Astronomy, The Ohio State University, 140 West 18th Avenue, Columbus OH 43210}
\altaffiltext{2}{Center for Cosmology and AstroParticle Physics, The Ohio State University, 191 W. Woodruff Avenue, Columbus OH 43210}
\altaffiltext{3}{Department of Physics and Astronomy, University of Oklahoma, 440 W. Brooks Street, Norman, OK 73019}

\begin{abstract}
We use the progenitor of SN~2012aw to illustrate the consequences of 
modeling circumstellar dust using Galactic (interstellar) extinction 
laws that (1) ignore dust emission in the near-IR and beyond;
(2) average over dust compositions, and 
(3) mis-characterize the optical/UV absorption by assuming that 
scattered photons are lost to the observer.
The primary consequences for the progenitor of
SN~2012aw are that both the luminosity and the absorption are 
significantly over-estimated.  In particular, the stellar 
luminosity is most likely in the range $10^{4.8} < L_*/L_\odot < 10^{5.0}$
and the star was not extremely massive for a Type~IIP progenitor,
with $M_* < 15M_\odot$.  Given the properties of the circumstellar
dust and the early X-ray/radio detections of SN~2012aw, the star
was probably obscured by an on-going wind with $\dot{M}\sim 10^{-5.5}$
to $10^{-5.0} M_\odot$/year at the time of the explosion, roughly consistent
with the expected mass loss rates for a star of its temperature ($T_* \simeq 3600_{-200}^{+300}$~K)
and luminosity.  In the spirit of Galactic extinction laws, we supply 
simple interpolation formulas for circumstellar extinction by dusty 
graphitic and silicate shells as a function of wavelength 
($\lambda \geq 0.3\mu$m) and total (absorption plus scattering) 
V-band optical depth ($\tau_V \leq 20$).  These do not
include the contributions of dust emission, but provide
a simple, physical alternative to incorrectly using 
interstellar extinction laws.
\end{abstract}

\keywords{stars: evolution -- supergiants -- supernovae:general}

\section{Introduction}
\label{sec:introduction}

A key component to understanding supernovae (SNe) is the
mapping between the explosions and their progenitor stars.  Slow,
steady progress is being made, and it is now well established that
Type~IIP SN are associated with red supergiants (see the review by
\citealt{Smartt2009}).  There is a puzzle, however, in that the 
observed upper limit of $\sim 16 M_\odot$ on the masses Type~IIP
SN progenitors appears to be significantly lower than the maximum masses 
of $\sim 25M_\odot$ for  stars expected to explode while still red supergiants (\citealt{Smartt2009b}),
part of a more general absence of massive SN progenitors (see \citealt{Kochanek2008}).
Since these ``missing'' progenitors should be more luminous than
those which are being discovered, they must either be hidden,
evolve differently than expected, or fail to explode.  For example,
in the rotating models of \cite{Ekstrom2012}, the upper mass limit
for red supergiants is lower, with $20M_\odot$ stars being blue
rather than red at the onset of carbon burning.  Alternatively, 
\cite{Oconnor2011} and \cite{Ugliano2012} find that the progenitor 
masses of $20$-$25M_\odot$ corresponding to the upper mass 
range for red supergiants are more prone to failed explosions and
prompt black hole formation, and such events would have to be found
by searching for stars disappearing rather than explosions 
appearing (\citealt{Kochanek2008}).  Binary evolution can also
alter the distribution of final states at the time of explosion
given the high probability for mass transfer as stars expand to
become red supergiants (e.g. \citealt{Sana2012}).

If the discrepancy is not explained by the physics of stellar evolution
or explosion, then the most likely remaining explanation is the effect
of circumstellar dust.  For example,
\cite{Walmswell2012} note that more massive and luminous red 
supergiants have stronger winds which can form dust and partly
obscure the star.  This then biases (in particular) the upper
mass limits associated with failed searches for progenitor 
stars, since the luminosity limits must be corrected
for an unknown amount of circumstellar
dust extinction.  \cite{Fraser2012} and \cite{Vandyk2012} recently
analyzed the progenitor of SN~2012aw, finding that it was both
relatively high mass ($15$-$20M_\odot$ for \cite{Vandyk2012} and
$14$-$26M_\odot$ for \cite{Fraser2012}) and the most heavily
obscured of any SN progenitor other than the completely
obscured (and debated) SN~2008S class (see \citealt{Prieto2008},
\citealt{Thompson2009}, \citealt{Kochanek2011}).  Since most
of the extinction vanished after the SN, the dust must have
been circumstellar (\citealt{Fraser2012}, \citealt{Vandyk2012}).

Like most studies of circumstellar dust in supernovae or supernovae
progenitors, \cite{Walmswell2012}, \cite{Fraser2012} and \cite{Vandyk2012}
treat circumstellar dust as if it is a foreground screen
that can be quantitatively modeled using  Galactic
interstellar extinction curve models parametrized by the value of $R_V$ 
(e.g. \citealt{Cardelli1989}).
Usually only studies that are self-consistently calculating
emission by the dust correctly model the absorption
by circumstellar dust (e.g. studies of the SN~2008S class of
transients by \citealt{Wesson2010}, \citealt{Kochanek2011}, 
or \citealt{Szczygiel2012}).  But three well-known effects
mean that it is never appropriate to make this approximation
unless the optical depth is negligible compared to the required
precision of the analysis.    First, emission
from circumstellar dust can be important in the near-IR if
the star is presently forming dust at temperatures of 
$1000$-$2000$~K.  This matters most for hotter stars with
less intrinsic near-IR emission than the
red supergiants we consider here.  Second, interstellar dust
has the average composition of dust from all sources, while
individual stars have the dust associated with their
particular chemistry. This is relevant here because 
$\sim 20 M_\odot$ stars generally produce silicate rather
than graphitic dusts (e.g. \citealt{Verhoelst2009}).

\begin{figure*}
\plotone{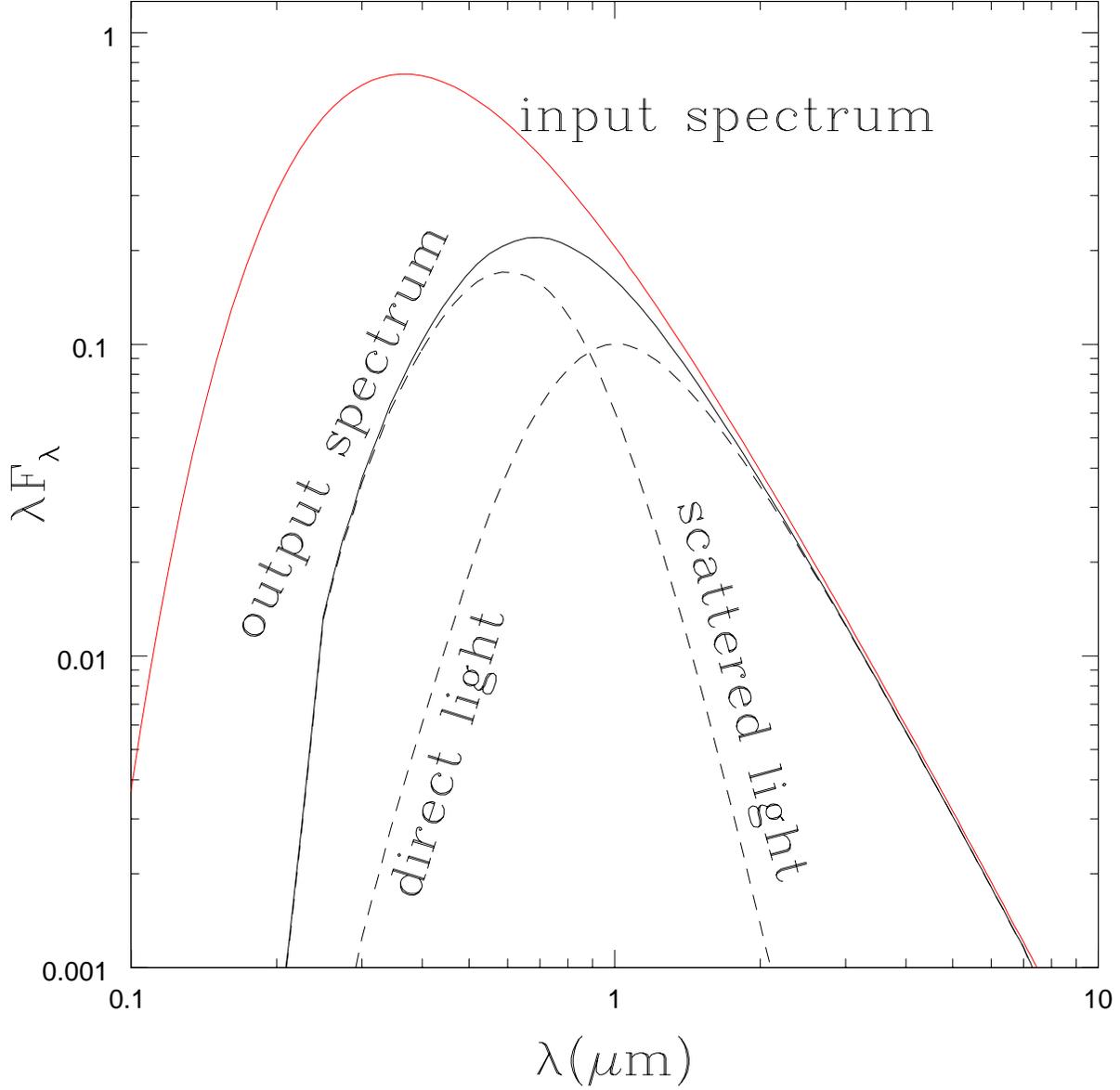}
\caption{
  DUSTY model of a $10^4$~K black body surrounded by $\tau_V=3$
  of cold silicate dust in a shell with $R_{out}/R_{in}=2$
  and $\rho \propto 1/r^2$ in the shell.  The dust converts the input spectral
  energy distribution (SED) to the 
  output SED, where the output SED is comprised of contributions
  from photons that escape without scattering (direct light) and
  photons that escape after being scattered (scattered light).  In an image,
  the direct emission is a point source and the scattered
  emission is a halo with a radius comparable to the inner
  radius of the shell. If the dusty
  shell is large enough to be resolved, only the direct
  emission is measured as coming from the source.
  We have made the dust cold enough to have no 
  contributions from dust emission over this wavelength range.
  }
\label{fig:modsed}
\end{figure*}

\begin{figure*}
\plotone{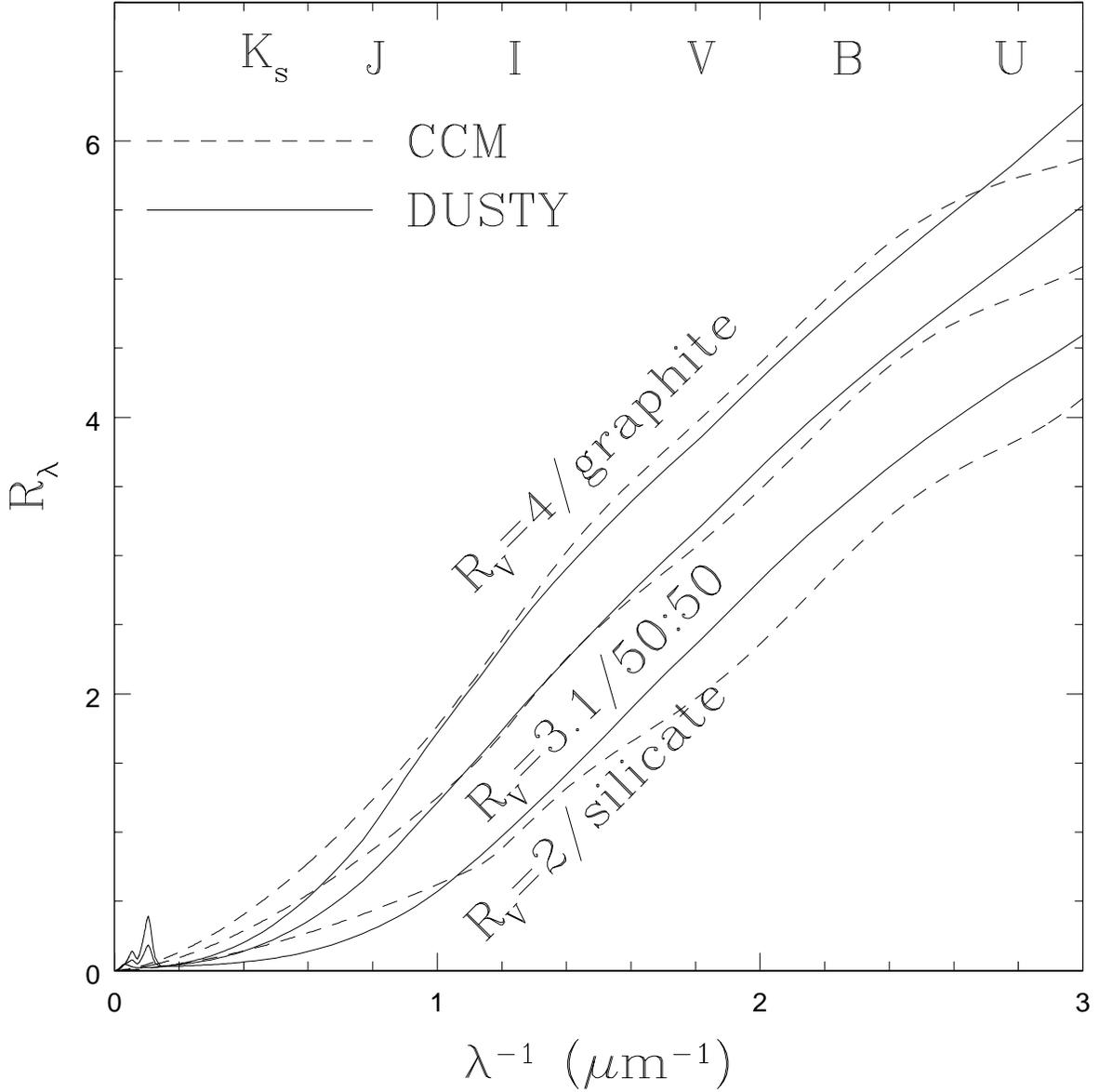}
\caption{
  Effective extinction laws $R_\lambda$ 
  if the dust is a foreground screen where we only measure the
  unscattered direct light.  The solid curves show the results for $\tau_V=3$
  of graphitic dust (top), silicate dust (bottom) and a 50:50 mix (middle).
  The dashed curves show the \cite{Cardelli1989} extinction laws (CCM) with 
  $R_V=4$ (top), $3.1$ (middle) and $2.0$ (bottom).  Normal
  Galactic dust ($R_V=3.1$) is relatively well modeled by 
  the 50:50 mix. 
  }
\label{fig:moddir}
\end{figure*}

\begin{figure*}
\plotone{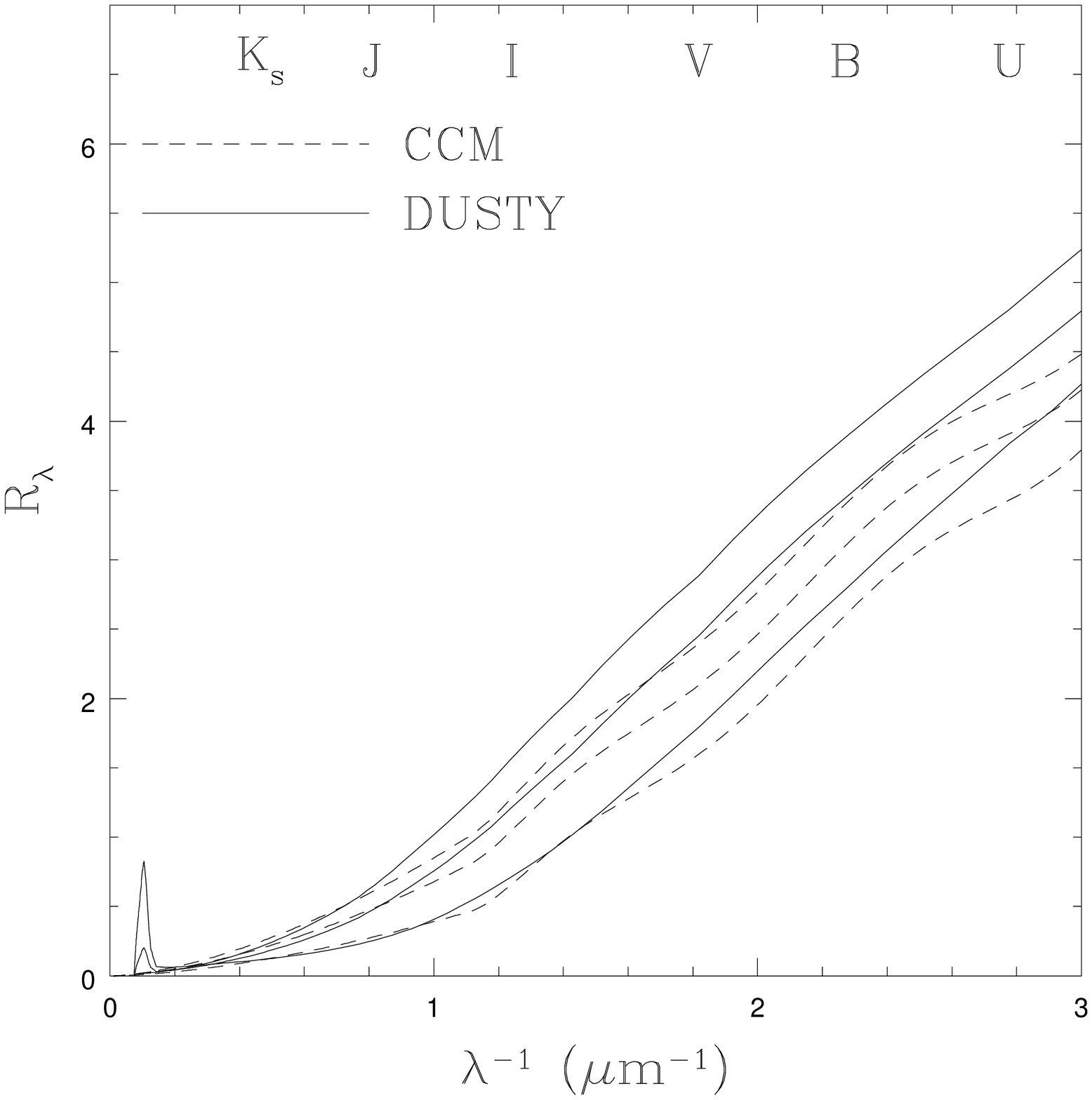}
\caption{
  Effective extinction laws $R_\lambda$ for dust in an unresolved shell around
  the star where we measure both the direct and scattered light.
  The solid curves show the results for $\tau_V=3$ of
  graphitic dust (top), silicate dust (bottom) and a 50:50 mix (middle).
  The dashed curves show the best fit CCM (\citealt{Cardelli1989})
  extinction laws with $R_V=2.4$ (top, graphitic), $R_V=2.1$
  (middle, 50:50 mix) and $R_V=1.6$ (bottom, silicate).
  The fits are never very good, particularly at high optical depth,
  and the best fit parameters vary somewhat with the 
  optical depth and wavelength.  
   \cite{Goobar2008} obtained 
  similar results.  The peak at
  long wavelengths (small $\lambda^{-1}$) is a silicate emission feature which
  DUSTY includes in the scattered light contribution to the 
  emerging spectrum.
  }
\label{fig:modall}
\end{figure*}

The third, and least appreciated point, is the different role
of scattered photons in interstellar and circumstellar extinction.
This issue has been discussed in detail by \cite{Wang2005} and
\cite{Goobar2008} in the context of anomalously low estimates of
$R_V$ for Type~Ia SN, but correctly treating the problem has not
become a matter of practice.  For a foreground screen, scattered
light forms a very diffuse, extended halo around the source 
which is not included in the estimate of the source flux.
This halo can sometimes be seen as the extended dust echoes
of transient sources (e.g. \citealt{Sugerman2002}).  If the 
circumstellar dust is unresolved, however, the scattered light is 
simply included in the total flux.  Thus, for interstellar 
extinction you observe only the direct, unabsorbed and unscattered
emission, while for circumstellar extinction you observe both the
direct unabsorbed and the scattered emission. {\it
Most of the observed optical emission from a moderately
self-obscured star is scattered light.}  

Fig.~\ref{fig:modsed} shows an example generated by
DUSTY (\citealt{Ivezic1997}, \citealt{Ivezic1999}, \citealt{Elitzur2001}) 
using the spectral
energy distribution (SED) of a $10^4$~K black body surrounded
by $\tau_V=3$ (scattering plus absorption) of cold silicate dust.  The dusty material has
a density profile of $\rho \propto 1/r^2$ in a shell with
$R_{out}/R_{in}=2$.  Galactic (interstellar)
extinction corresponds to putting the dust at such
a large radius that we no longer include the scattered
light in the observed flux of the star, and the
extinction law corresponds to the wavelength dependent
difference between the input spectrum and the directly
escaping spectrum.  Fig.~\ref{fig:moddir} shows this 
ratio converted into an extinction law $R_\lambda$ for
pure graphitic, pure silicate and a 50:50 mix of the
two, as compared to \cite{Cardelli1989} Galactic extinction
laws with $R_V=4.0$, $3.1$ and $2.0$.  Galactic dust
models typically have a roughly 50:50 mix of graphitic
and silicate grains (see the summary in \citealt{Draine2011}), and 
using this mix in the DUSTY
models roughly reproduces a typical $R_V=3.1$ Galactic
extinction law.  The match is not perfect because there
are differences in the assumed size distributions.  

For a star surrounded by an unresolved
shell of dust, however, the emission we measure is the
sum of the direct and scattered light, which is very
different from the direct emission alone.  Fig.~\ref{fig:modall}
shows this case converted into an effective circumstellar extinction
law for the same three dust mixtures.  For circumstellar
dust, all three examples are now significantly below
the $R_V=3.1$ curve and a given change in color implies
a significantly smaller change in luminosity.    
If we fit the absorption in these DUSTY models with \cite{Cardelli1989}
extinction models, the best fits for the graphitic,
silicate and 50:50 dusts are $R_V\simeq 2.4$, 
$R_V\simeq 1.6$ and $R_V \simeq 2.1$, respectively. These
are the best fits for $\tau_V=3$ from $0.36\mu$m (U band)
to $1\mu$m.  The best fits vary modestly ($\Delta R_V \simeq 0.1$)
with $\tau_V$ and the fitted wavelength range, become worse
for higher optical depths, and there are no truly good fits.
Fig.~\ref{fig:modall} superposes these \cite{Cardelli1989} extinction curves
on the DUSTY models.  \cite{Goobar2008} found similar
estimates for the best \cite{Cardelli1989} extinction curves
to model circumstellar dust and also noted that they are 
not very good approximations.  

The final point to note in Figs.~\ref{fig:moddir} and \ref{fig:modall}
is that the dust composition is
quantitatively important.  Interstellar dust is a mixture
of graphitic and silicate dusts from many sources. 
Individual stars, however, typically only produce silicate
or graphitic dusts depending on the carbon to oxygen 
abundance ratio of the stellar atmosphere.  To zeroth
order, all available carbon and oxygen bond to make CO.
Then, a carbon poor (rich) atmosphere has excess oxygen
(carbon) to make silicate (graphitic) dusts.  The atmospheres 
of $\sim 20M_\odot$ red supergiants usually form silicate
dusts (e.g. \citealt{Verhoelst2009}) because they never
undergo the dredge up phases that enrich the atmospheres
of the lower mass AGB stars with carbon (e.g. \citealt{Iben1983}).
Assuming a mixed interstellar composition
will generally overestimate the absorption associated with
a given change in color for silicate dusts and underestimate
it for graphitic dusts.

In this paper we consider the effects of dust on the progenitor
of SN~2012aw in more detail.  First, in \S\ref{sec:midir} we 
use archival Spitzer IRAC data to set stringent limits
on the 3.6, 4.5, 5.8 and $8.0\mu$m fluxes of the progenitor,
and then model the spectral energy
distribution (SED) of the progenitor with DUSTY in order to
correctly model absorption, scattering and emission from a 
dust enshrouded star.  We highlight where any extinction
law derived for dust screens at great distances from the source
such as the standard \cite{Cardelli1989} extinction laws can create problems.
Finally, in \S\ref{sec:discussion} we summarize the results and
provide simple interpolation formulas for extinction by
graphitic and silicate circumstellar dust as a function
of wavelength and optical depth.

\begin{figure*}
\plotone{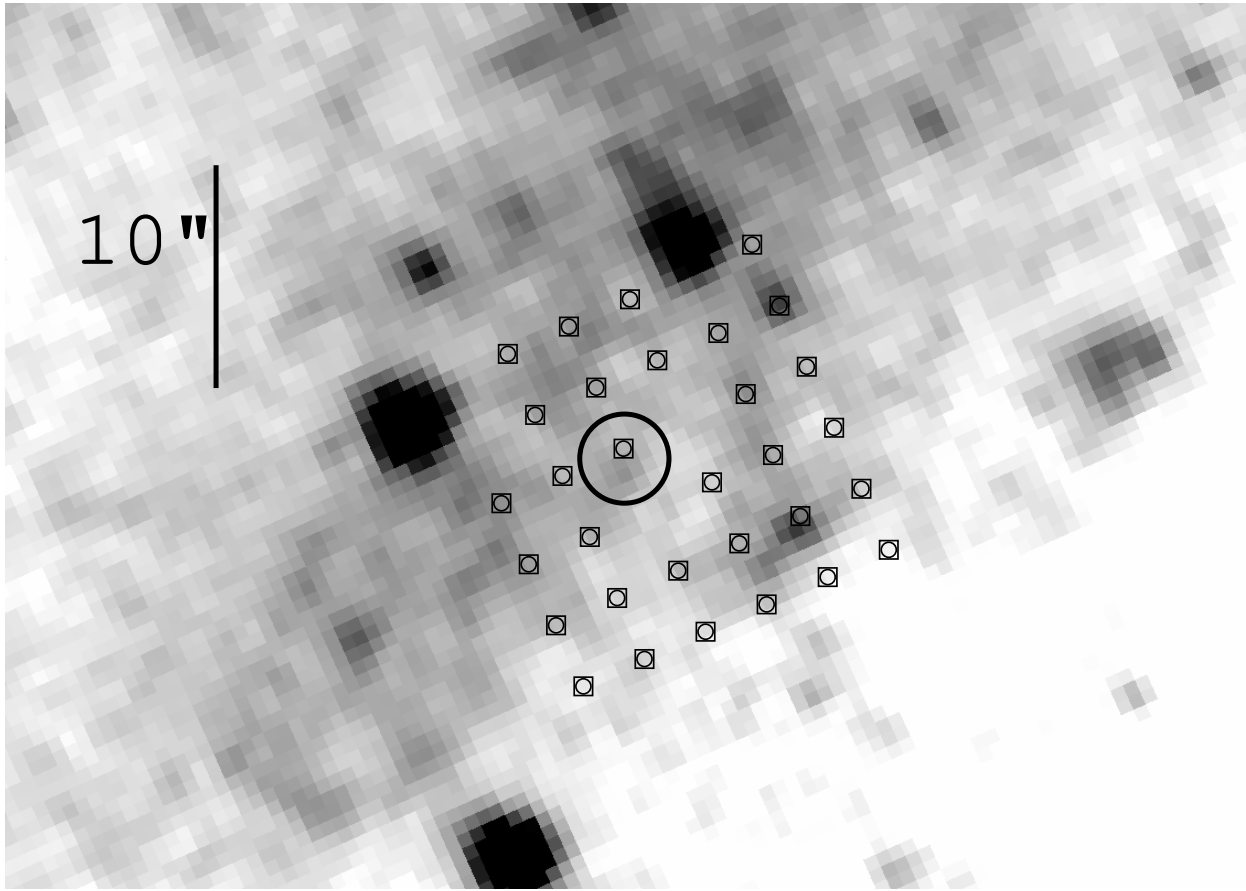}
\caption{
  Combined $3.6$ and $4.5\mu$m images of the region surrounding the
  progenitor of SN~2012aw.  The 2\farcs0 diameter circle marks the
  estimated position of the progenitor, although the formal uncertainties
  in the position are far smaller ($0\farcs05$).  The smaller points
  show the grid of apertures used to estimate the flux limits, where
  missing points show the locations of dropped apertures (see text).  
  A few of the grid points lie near fainter sources that are obvious
  in this co-added image but only marginally detectable (at about
  $2\sigma$) in the four individual images combined to make Fig.~\ref{fig:image}.
  The image is aligned to the world coordinate system with North
  up and East left.
  }
\label{fig:image}
\end{figure*}

\section{Mid-IR Flux Limits, X-ray Fluxes and Models }
\label{sec:midir}

As noted by \cite{Fraser2012}, there are no obvious sources in the
archival SINGS (\citealt{Kennicutt2003}) IRAC images of the region around the progenitor.  It is,
however, a relatively clean region with the emission dominated by
unresolved stars, as shown in Fig.~\ref{fig:image}.
 We  identified 8 nearby stars in common between 
the HST I-band image used by \cite{Fraser2012} and \cite{Vandyk2012}
to identify the progenitor and the image formed by co-adding 
the $3.6$ and $4.5\mu$m data and used the IRAF geomap/geotran
scripts to estimate the position of the progenitor shown in
Fig.~\ref{fig:image}.  Based on jackknife resampling of the
reference stars, the nominal uncertainties in the position
are small (about 0\farcs05).

Since we are essentially confusion limited, we estimated flux limits 
using the grid of 36 apertures covering the location of the progenitor
and bounded by the nearby brighter stars as shown in Fig.~\ref{fig:image}.  
The grid spacing was 3\farcs0.  We used the 
IRAF\footnote{IRAF is distributed by the 
  National Optical Astronomy Observatory, which is operated by the Association of 
  Universities for Research in Astronomy (AURA) under cooperative agreement with 
  the National Science Foundation.} 
APPHOT/PHOT to measure aperture magnitudes with a signal radius of 2\farcs4
and a sky annulus from 2\farcs4 to 7\farcs2.  The fluxes were calibrated 
following the procedures given in the \textit{Spitzer} Data Analysis  
Cookbook\footnote{http://irsa.ipac.caltech.edu/data/SPITZER/docs/dataanalysistools/}
and aperture corrections of 1.213, 1.234, 1.379, and 1.584 for the
3.6 through $8.0\mu$m bands.  Keeping the sky annulus immediately next to the 
signal aperture minimizes the effects of background variations created by the
galaxy.  We used a $2\sigma$ outlier rejection procedure in 
order to exclude sources located in the local sky annulus, and correct for the 
excluded pixels assuming a Gaussian background flux distribution.

We clipped 3 of the grid points in the
wings of the brighter sources and the two lowest (negative) flux points 
that lay in an obvious ``dark lane'', leaving 31 points.  Based on one
of the two SINGS epochs, the variances in the four bands are $10.7$, 
$11.4$, $12.8$ and $14.2$~$\mu$Jy for the $3.6$, $4.5$, $5.8$, and $8.0\mu$m 
bands respectively.  This is well above the nominal noise level for
the exposure time ($0.5$, $0.8$, $6$, and $6$~$\mu$Jy) because there
is stellar emission in the region.  The variance after coadding the two 
SINGS epochs is only slightly smaller, as expected for noise dominated by
confusion, so we rather conservatively use the variances 
derived from the single epoch as $1\sigma$ flux limits.  
Note that some of the grid points clearly lie on sources 
in the coadded image.  These sources are
not readily apparent in the four individual sub-images, where their detection
significance would be roughly $2\sigma$.  A source of similar flux at 
the supernova position would be equally obvious in Fig.~\ref{fig:image}.

For models of the spectral energy distribution, we combined the
V and I band estimates from  \cite{Vandyk2012} and \cite{Fraser2012},
including the difference in their V-band estimates as an additional
systematic error.  We used the J and K$_s$ estimates from \cite{Vandyk2012},
based on their better photometric calibrations.  We use a minimum
flux uncertainty of 20\%, which broadens the V, I and J-band
photometric errors, given that the data were obtained over a
long time period (1994-2009, with the SINGS IRAC data being obtained 
on 22/23 May 2004) and \cite{Vandyk2012} argue for the detection
of variability at V and I.   We model the star using the 
MARCS (\citealt{Gustafsson2008}) stellar atmosphere models, considering
only the $15 M_\odot$, solar metallicity, $5$~km/s turbulent velocity
$\log_{10} g = -0.5$ models preferred by \cite{Vandyk2012} and also considered by \cite{Fraser2012}.
These are available for effective temperatures of $T_e=3300$
to $4000$~K in steps of $100$~K and then $4250$ and $4400$~K.  
For intermediate temperatures we linearly interpolate between the available
models.  We averaged the high resolution MARCS models and the filter
transmission functions onto a coarser wavelength grid for use in
the DUSTY models and then generated magnitude estimates for each model
including the appropriate averages over the filter band passes. 
We first fit the measurements to
normalize the luminosity, and then added a contribution from the upper 
limits as $\Delta \chi^2 = (F_{model}/F_{limit})^2$ for each band,
where $F_{model}$ is the estimate from the normalized model and $F_{limit}$
is the $1\sigma$ limit on the flux.   

\cite{Immler2012} reported a Swift (\citealt{Gehrels2004}) X-ray detection 
of the SN in the first
week and \cite{Stockdale2012} and \cite{Yadav2012} report $20$~GHz radio
detections in the first month.  Swift continued to monitor the SN for an
extended period, so as an independent probe of the circumstellar medium
we obtained all the archival Swift XRT data, binning the observations as
summarized in Table~\ref{tab:xray}.  We reprocessed the XRT data using the
\verb+xrtpipeline+ tool provided by the Swift team.  We binned the observations
into six roughly logarithmic time intervals and then
reprojected each group of observations into a single image. 
We chose the source region to be a circle
centered on the SN with a radius of 10 pixels (23\farcs6) and a nearby
background region without any sources, using aperture corrections based
on the analysis of the Swift PSF by \cite{Moretti2005}. Following
\cite{Immler2012} we assumed a Galactic foreground column density of
$2.9\times 10^{20}$ (\citealt{Dickey1990}).  For our epoch best matching
\cite{Immler2012} we reproduce their count rates.  While none of the
X-ray detections are of very high significance, there appears to be
a low level of detectable ($3\sigma$) and probably time varying
 X-ray emission for the first 1--2 months.
There are no clear detections at low energies ($0.2$--$0.5$~keV) and
only very marginal detections at high energies ($2$--$10$~keV) -- the
observed counts are completely dominated by the $0.5$--$2$~keV band.

We use the DUSTY (\citealt{Ivezic1997}, \citealt{Ivezic1999}, \citealt{Elitzur2001})
dust radiation transfer models to correctly include dust emission, absorption,
scattering and composition when modeling the progenitor SED.  We use either graphitic or silicate dust models from
\cite{Draine1984} and the default \cite{Mathis1977} power-law
grain size distribution ($dn/da \propto a^{-3.5}$ for $0.005\mu\hbox{m}<a<0.25\mu\hbox{m}$).
The dust lies in a spherical shell with a $1/r^2$
density distribution and a $R_{out}/R_{in}=10$ axis ratio between the inner
and outer radii.  While the directly escaping light depends only on the optical 
depth, thicker shells (larger $R_{out}/R_{in}$) allow more scattered light to    
escape than thin shells for the same total optical depth, but this is a second
order effect and we will hold $R_{out}/R_{in}$ fixed.
The model parameters are the luminosity of the star, $L_*$, the temperature of the 
star, $T_*$, the temperature $T_d$ of the dust at $R_{in}$, and the dust optical 
depth $\tau_V$ at V-band ($0.55\mu$m).  This is the total (absorption plus
scattering) optical depth, $\tau_{tot} = \tau_{abs}+ \tau_{scat}$, where the scattering
optical depth is related to the total optical depth by the albedo, $\tau_{scat}=w \tau_{tot}$.    
The effective absorption optical depth is approximately 
$\tau_e = (\tau_{tot}\tau_{abs})^{1/2}=(1-w)^{1/2} \tau_{tot}$.  At V-band,
the standard DUSTY silicate (graphitic) dust models have albedos of 
$w_V=0.86$ ($0.47$), so for the same effective absorption the silicate models
require significantly higher total optical depths $\tau_V$.  
The V-band extinction is approximately $A_V = 2.5\tau_{e,V}/\ln 10 = 0.41\tau_V$ ($0.79\tau_V$)
for the silicate (graphitic) model.  The radius $R_{in}$ is then a 
derived quantity given the model parameters.  We fit the data and
estimate uncertainties by embedding DUSTY in a Markov Chain Monte Carlo 
(MCMC) driver.  Without mid-IR detections, we cannot determine the dust
temperature, so we simply considered cases with $T_d=100$~K, $500$~K,
$1000$~K and $1500$~K.  We assume a fixed foreground Galactic extinction
of $E(B-V)=0.05$~mag (modeled using an $R_V=3.1$ extinction law) and a 
distance modulus of $\mu=30$~mag (\citealt{Freedman2001}).  

While they are not physically correct models, we do recover the results of
\cite{Vandyk2012} and \cite{Fraser2012} if we model our version of the SED
including the mid-IR flux limits but no dust emission and 
using \cite{Cardelli1989} extinction laws.
Like \cite{Vandyk2012} we find that higher $R_V$ extinction laws provide
better fits, but we also find a stellar temperature degeneracy similar
to \cite{Fraser2012}.  Clearly the models have a ``degeneracy''
direction which is very sensitive to the exact assumptions of the
model and can lead to either tight or loose limits on the effective
temperature.  The specific rationale for using $R_V\simeq 4$ extinction
laws in \cite{Vandyk2012} is not really valid,\footnote{
  \cite{Vandyk2012} advocate a high $R_V$ based on the finding
  of \cite{Massey2005} that matching spectroscopic and B$-$V
  color estimates of $A_V$ requires using larger effective
  value of $R_V$ for the photometric estimate.  However, the problem
  \cite{Massey2005} was addressing does not correspond to changing
  $R_V$ at all wavelengths in a general fit to an SED.
   \cite{Massey2005} were correcting for the shift in effective
   filter wavelengths at blue and ultraviolet wavelengths between
   the O stars usually used to build models for the extinction curve
   and low temperature red supergiants.  The effective wavelength
   of a filter is an average of the wavelength weighted by the
   spectrum over the filter transmission function.  Where the
   spectrum is changing rapidly and cannot be modeled as a linear
   trend, the effective wavelength shifts.  This is primarily
   an issue on the ``Wien'' tails of stellar spectra, where the
   effective wavelength for a cool star is significantly redder
   than that of a hot star.  For the $3600$~K MARCS models, the
   B, V and I band shifts relative to an O star are roughly 8\%,
   3\% and 2\%, and then become steadily smaller at longer
   wavelengths.  \cite{Massey2005} derive an effective
   correction for converting observed B$-$V colors to
   V-band extinctions ($A_V$).  Mathematically,
   the observed color is $(B-V)_{obs}=(R'_B-R'_V)E(B-V)$,
   where $R'_B$ and $R'_V$ are the $R_V=3.1$ extinction
   law at the shifted, effective wavelengths of the B and
   V bands for the cold star.   Thus, getting the correct $A_V$ from
   the observed color corresponds to using $R_V^e=3.1/(R'_B-R'_V)$
   so that $A_V=R_V^e (B-V)_{obs}$.  For the $3600$~K MARCS models,
   the effective B and V-band wavelength shifts then lead to the
   $R_V^e\simeq 4.1$ values reported by \cite{Massey2005}.
   {\it This is a specific correction for the distortion of the
   B$-$V colors by the wavelength shifts and not a change in
   the overall extinction curve.}   At redder wavelengths, the
   wavelength shifts become steadily smaller and the differences
   between the true and effective (band-averaged) extinction
   curves become smaller and smaller.  
    Since the B and U band fluxes only enter as weak upper 
    limits on the progenitor, it is likely more correct to simply ignore
   the shifts at V band and longer wavelengths than to assume that an underlying
   $R_V=3.1$ extinction curve is globally transformed into a $R_V \simeq 4.35$
   extinction curve.
  }
but allowing freedom in the extinction curve is certainly a
more conservative approach.  
For example, if we fix $R_V=4.35$
and $T_*=3600$~K to match \cite{Vandyk2012}, we obtain a 
luminosity of $\log_{10} L_*/L_\odot = 5.15 \pm 0.05$ compared
to $5.21\pm0.03$. If we fix $R_V=3.1$ to match \cite{Fraser2012},
then we find $\log_{10} L_*/L_\odot = 5.10\pm0.05$ for $T_*=3600$~K
and $\log_{10} L_*/L_\odot = 5.43\pm0.05$ for $T_*=4400$~K,
close the values of $\log_{10} L_*/L_\odot = 5.0$ and $5.6$ they find for temperatures
of $T_*=3550$~K and $4450$~K using MARCS models with $\log_{10} g=0$. 

\begin{figure*}
\epsscale{0.55}
\plotone{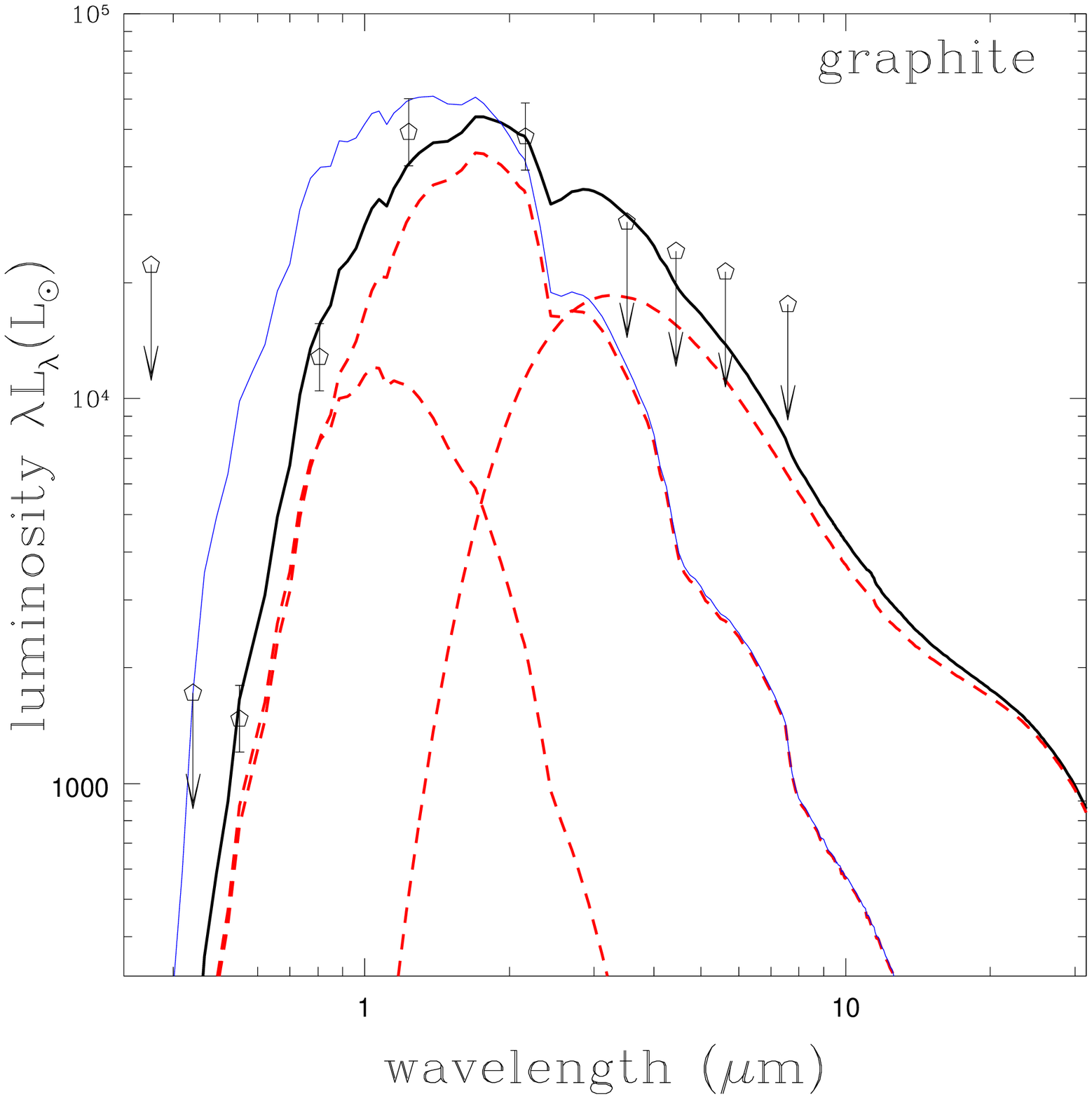}
\epsscale{0.55}
\plotone{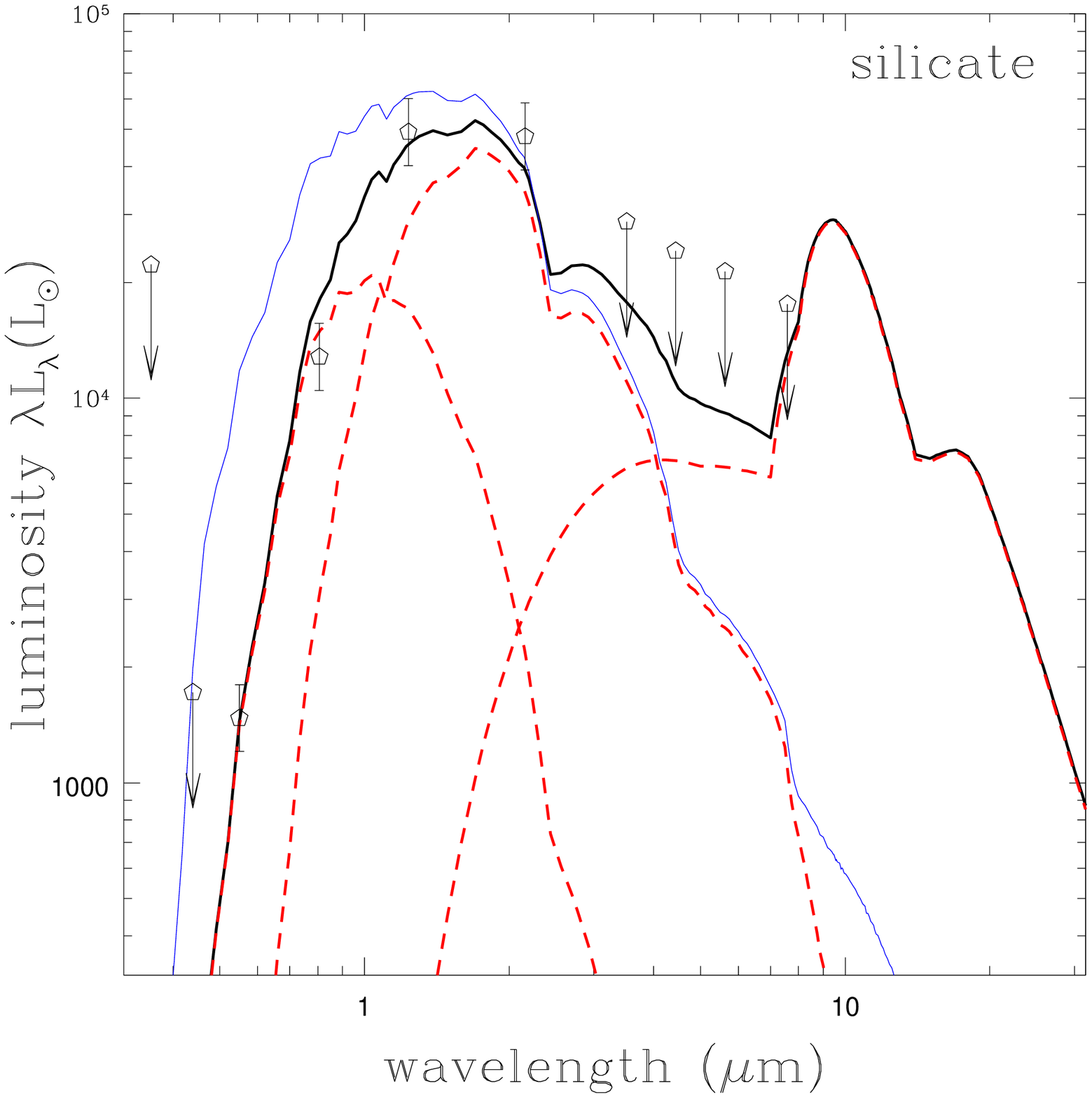}
\epsscale{1.0}
\caption{
  Graphitic (top) and silicate (bottom) fits to the SED of
  the progenitor for $T_d=1000$~K at the inner edge of the
  dust distribution.  The heavy black curves show the 
  SED model, which is comprised of (from left to right)
  scattered, direct and dust emission, as shown by the
  dashed red curves.  The thin blue curves show 
  the model for the
  unobscured SED of the star.  The open symbols show the 
  measured luminosities and limits.  Note the different
  balance between absorption and scattering for the two
  dust compositions and the modest contribution of dust
  emission to the K$_s$ band.  
  }
\label{fig:sed}
\end{figure*}

\begin{figure*}
\plotone{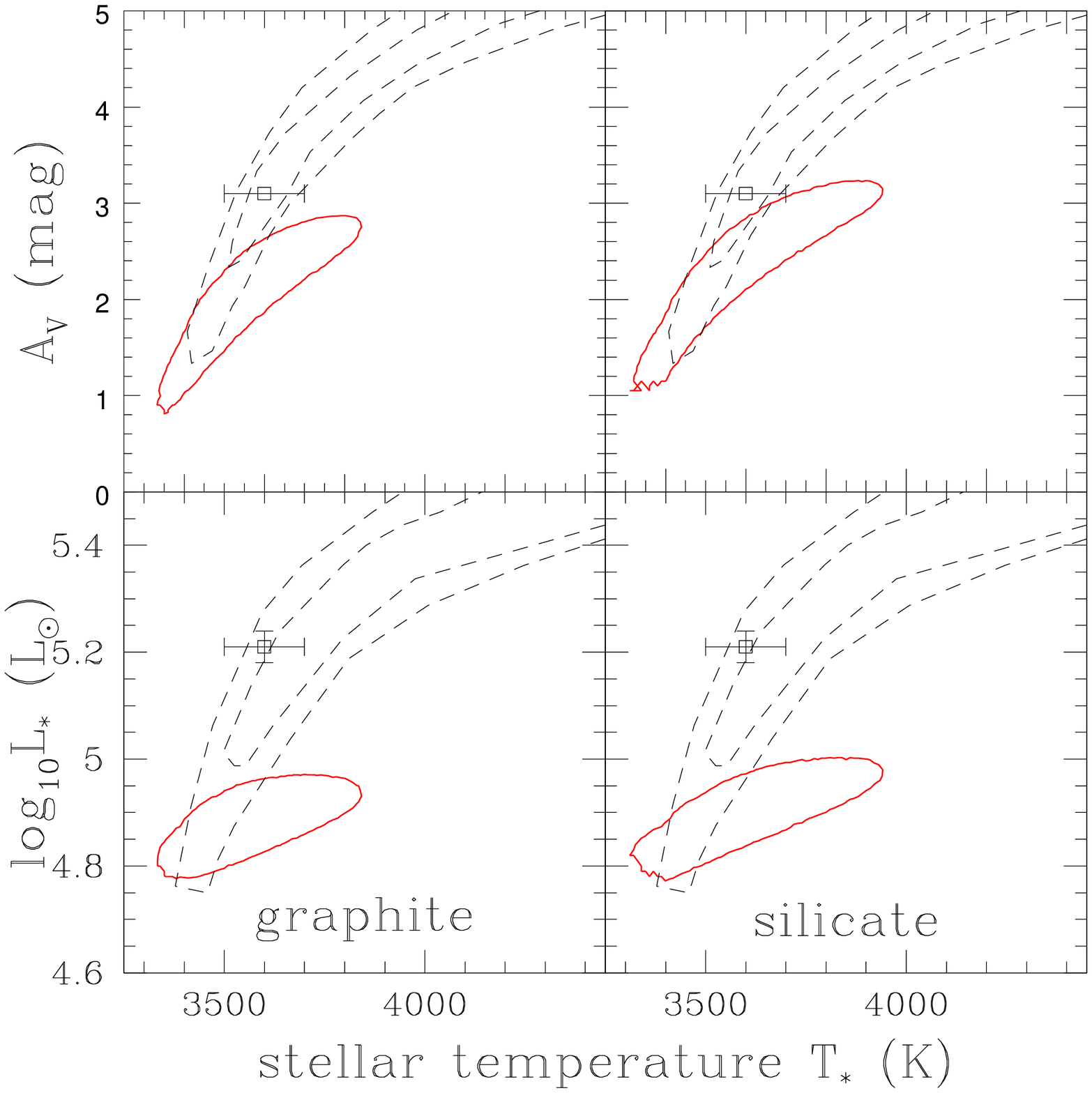}
\caption{
  Constraints on the visual extinction $A_V$ (top) and stellar luminosity 
  $L_*$ (bottom) as a function of stellar temperature $T_*$.  The 
  solid contours show the $1\sigma$ constraints on two parameters
  ($\Delta\chi^2=2.3$) for the graphitic (left) and silicate (right)
  models and an inner edge dust temperature of $T_d=1000$~K roughly
  corresponding to having a dust forming wind at the time of the
  observations.  The points and dashed contours show the results 
  from \cite{Vandyk2012} and \cite{Fraser2012}, respectively.
  The DUSTY model $\tau_V$ were converted to 
  $A_V=2.5(1-w_V)^{1/2}\tau_V/\ln 10$.
  }
\label{fig:dmodels}
\end{figure*}


Fig.~\ref{fig:sed} shows two representative DUSTY models of the
progenitor for circumstellar
silicate and graphitic dusts with $T_d=1000$~K, roughly corresponding
to the expected inner edge dust temperatures if the star was
forming dust at the time of the observations.  Formally, the
graphitic model is a better fit ($\chi^2=5.5$ versus $8.8$),
but the models have essentially identical stellar luminosities, 
$L_* = 10^{4.9}L_\odot$, and temperatures, $T_* \simeq 3550$~K.
Both models marginally satisfy the mid-IR flux limits at this
luminosity, with the graphitic models primarily limited by
the non-detections at $3.6$ and $4.5\mu$m, and the silicate
models limited by the contribution of silicate emission peak 
at $8.0\mu$m.  

Fig.~\ref{fig:sed} illustrates all three of the basic points
about circumstellar dust as compared to interstellar dust.
 First, the optical flux is
completely dominated by the scattered emission that is
not included in interstellar extinction laws.  Second,
dust emission is quantitatively important to the K$_s$
band flux.  Because the star is cold, dust emission does
not dominate the near-IR flux as it would for a hot star
of the same luminosity and degree of obscuration, but 
it does partly compensate for the absorption.  Third,
there are quantitative differences between the two
dust types in their balance between absorption and
scattering and the nature of the mid-IR emission.
The total optical depths of the models are quite different,
$\tau_V=5.9$ for silicates and $\tau_V=2.6$ for graphite,
but the effective absorption optical depths of the two
models are very similar, with $\tau_{e,V}=(1-w_V)^{1/2}\tau_V=2.2$
for silicates and $1.9$ for graphite.

Fig.~\ref{fig:dmodels} shows the allowed parameter ranges for
the stellar luminosity $L_*$ and visual extinction $A_V$ as
a function of stellar temperature $T_*$ as compared to the
results from \cite{Fraser2012} and \cite{Vandyk2012}.  We
show the results for $T_d=1000$~K, which roughly corresponds
to the inner edge dust temperature if there was a dust forming
wind at the time of the observations.  Table~\ref{tab:sedmodels}
gives the results for the other dust temperatures.  They are
generally similar except for the $T_d=100$~K graphitic model.
The key issue for the progenitor is that the preferred solutions
of both \cite{Fraser2012} and \cite{Vandyk2012} significantly
overestimate the luminosity of the progenitor.  Our results
agree with \cite{Vandyk2012} on the temperature, but the luminosity
$L_*$ is a factor of two lower.  The mid-IR limits essentially
preclude the higher luminosity and temperature solutions of 
\cite{Fraser2012} unless the dust temperature is made 
significantly lower than $T_d=500$~K and the dust emission
moves out of the IRAC band passes.  Even there, only the 
cold $T_d=100$~K graphitic model allows a significantly higher
luminosity.  As expected, a large part of the difference is
that the models based on interstellar extinction are significantly
overestimating the overall absorption.

\begin{figure*}
\plotone{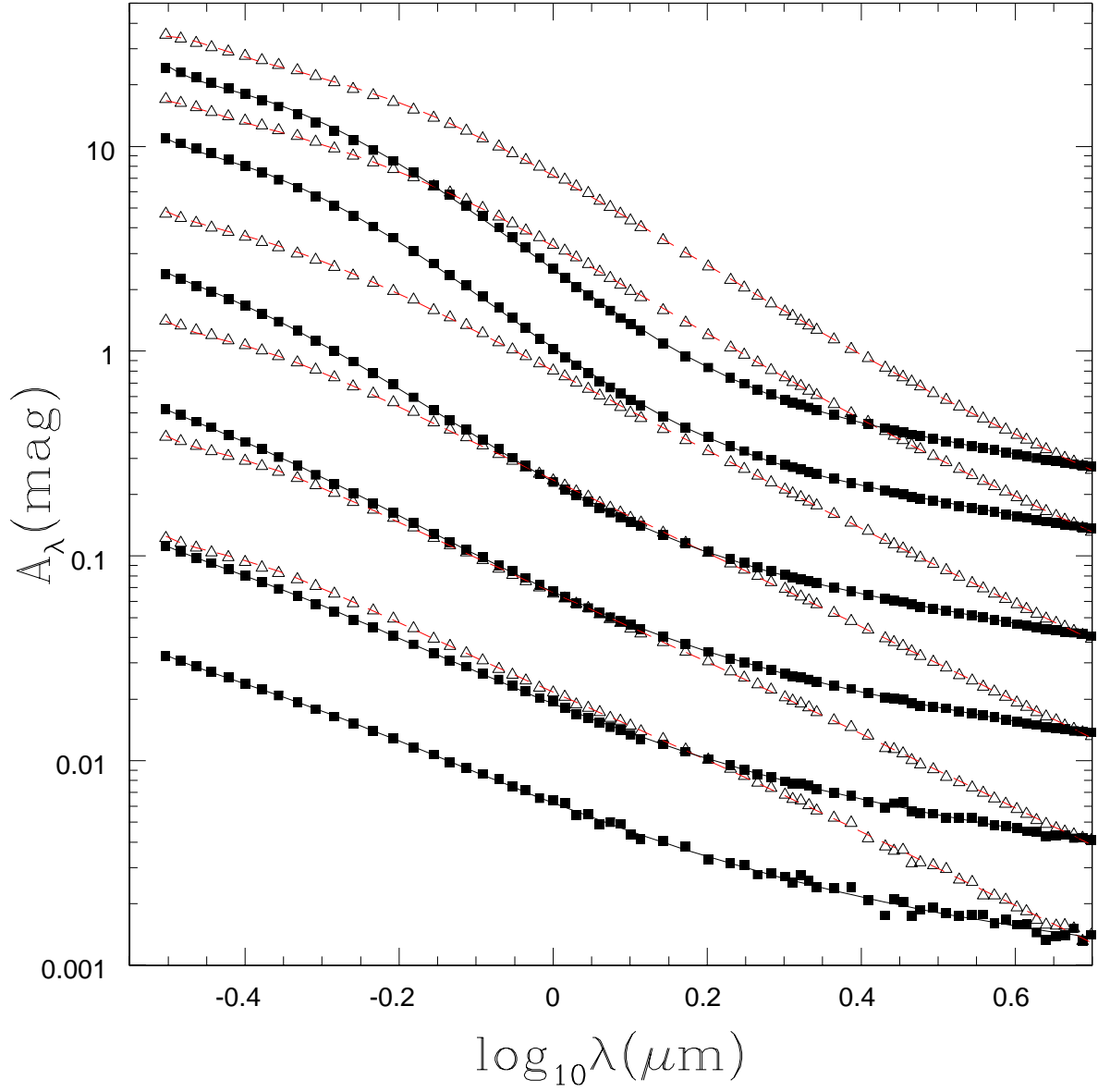}
\caption{
  Extinction $A_\lambda$ as a function of wavelength for silicate
  (squares) and graphitic (triangles) circumstellar dust with
  $R_{out}/R_{in}=10$ and optical
  depths of $\tau_V=0.1$ (bottom), $0.3$, $1.0$, $3.0$, $10.0$ and
  $20.0$ (top).  
    Note how the shapes of the curves depend on 
  both composition and optical depth.
   The curves through the points are the model from 
  Table~\ref{tab:models}.
  }
\label{fig:opdepth}
\end{figure*}

\begin{figure*}
\plotone{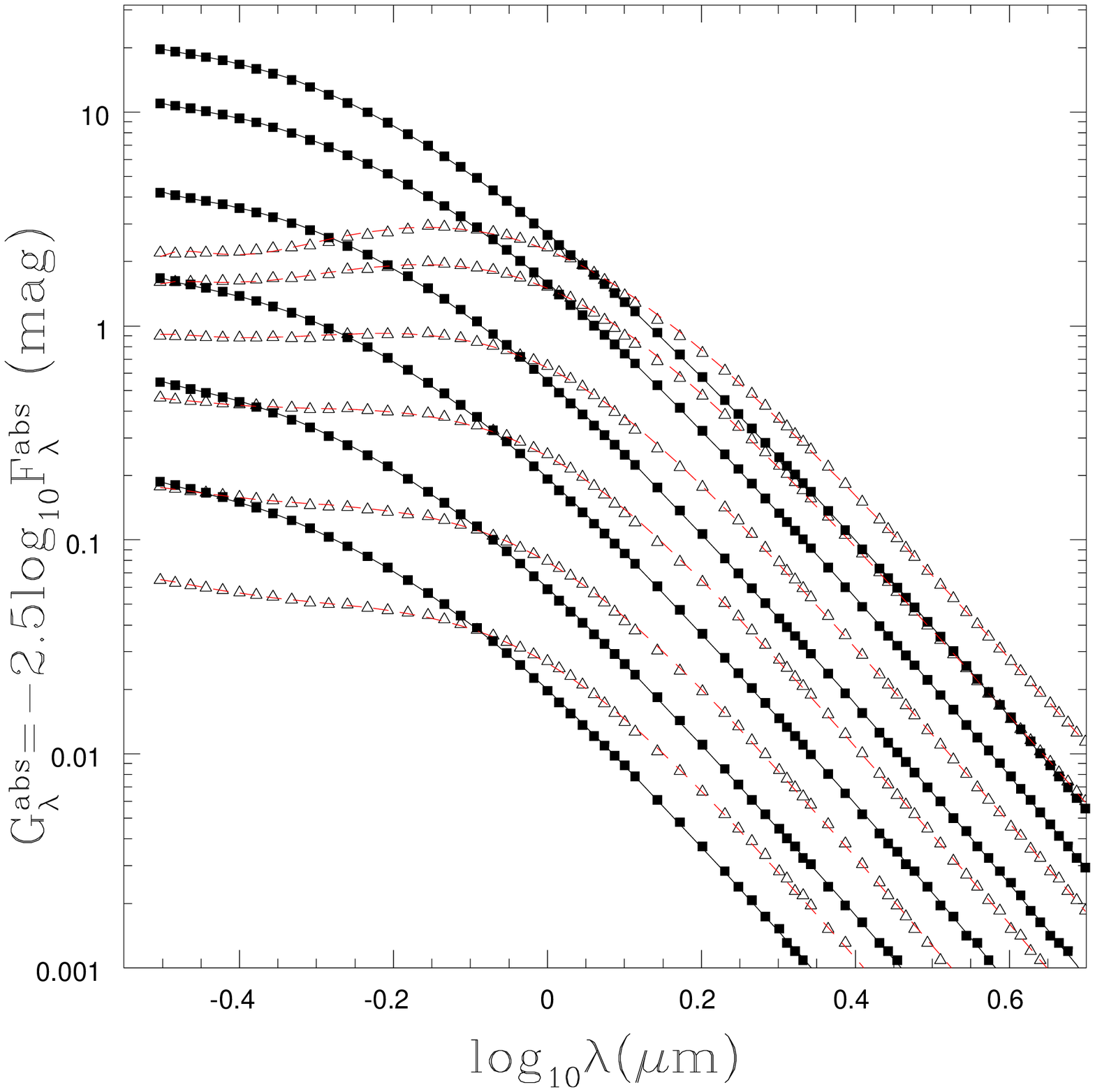}
\caption{
  Interpolation function $G_\lambda^{abs}=-2.5\log_{10}F_\lambda^{abs} $ for the fraction of the observed
  light that is direct emission as a function of wavelength for silicate
  (squares) and graphitic (triangles) circumstellar dust with
  $R_{out}/R_{in}=10$ and optical
  depths of $\tau_V=0.1$ (bottom), $0.3$, $1.0$, $3.0$, $10.0$ and
  $20.0$ (top).  The curves through the points are the model from 
  Table~\ref{tab:direct}.  For an input SED $S_\lambda$, the observed
  SED is $S_\lambda 10^{-0.4A_\lambda}$ of which 
  $S_\lambda 10^{-0.4A_\lambda} F_\lambda^{abs} =S_\lambda 10^{-0.4(A_\lambda+G_\lambda^{abs})}$
  is directly escaping emission and  $S_\lambda 10^{-0.4A_\lambda} (1-F_\lambda^{abs})$
  is scattered emission.
    As in Fig.~\ref{fig:opdepth}, note how the shapes of the curves depend on 
  both composition and optical depth.
  }
\label{fig:direct}
\end{figure*}

\section{Discussion}
\label{sec:discussion}

\cite{Fraser2012} and \cite{Vandyk2012} argue that the progenitor
was probably the most massive yet found for a Type~IIP SN, probably
at or above the upper limit of $(16.5\pm1.5)M_\odot$ 
\cite{Smartt2009b} found in their statistical analysis of the masses
of Type~IIP progenitors.  Given the amount of circumstellar extinction, 
this seemed to match the suggestion by \cite{Walmswell2012} that the
upper mass limit could be biased by increasing levels of dust formation
for the more massive red supergiants due to the increase in mass loss
rates with luminosity.  
Here we argue that many of these inferences are biased by incorrectly
modeling circumstellar dust with an interstellar extinction law, 
thereby overestimating both the amount of extinction and the luminosity
of the star.  When we model the SED using circumstellar dust models,
the luminosity of the star is $L_* < 10^5 L_\odot$ and the mass is 
$M_* < 15M_\odot$, where the downwards shifts are easily understood
from the differing physics of interstellar and circumstellar extinction.
The visual extinction is overestimated because interstellar extinction
neglects the contributions of scattered light,   and, to a lesser extent
because of the low stellar temperature, the near-IR (K$_s$) extinction is 
overestimated by neglecting the emission from hot dust.  The absence
of a mid-IR source noted by \cite{Fraser2012} is a key point of 
evidence, since there should have been a detectable source given the
proposed, higher luminosities unless the dust is cold and emitting
at longer wavelengths than the IRAC bands.  In this particular
case, the dust composition has little effect on the inferred 
properties of the star.   

Unfortunately, without measuring the mid-IR portion of the SED we 
cannot determine the dust temperature $T_d$ other than limiting it
to be lower than the dust destruction temperature ($T_d \simeq 1500$~K).
However, the wind properties are a strong function of the dust
temperature because at fixed optical depth far more mass is required 
if the material is at larger radii and colder temperatures.  Ignoring
the minor (10\%) corrections from the finite value of $R_{out}/R_{in}=10$,
the mass loss rate required to support the optical depth is  
\begin{equation}
   \dot{M} =  4 \pi R_{in} v_w \tau_V \kappa_V^{-1}
        \simeq 10^{-5} 
        R_{in15} v_{w10} \tau_{V5} \kappa_{V100}^{-1}
    M_\odot~\hbox{year}^{-1}
    \label{eqn:mdot}
\end{equation}
where $R_{in} = 10^{15} R_{in15}$~cm, $v_w=10 v_{w10}$~km/s,
$\tau_V = 5\tau_{V5}$ and $\kappa_V = 10^2 \kappa_{v100}$~cm$^2$/g.
Table~\ref{tab:sedmodels} gives the values for the individual
models.  Supporting $\tau_V \sim 5$ when $T_d \simeq 100$~K requires
mass loss rates of order $\dot{M} \sim 10^{-2.5}M_\odot$/year and
implies total wind masses of order $3 M_\odot$ that are implausible
for a star with an initial mass of $M_*<15M_\odot$. 
For comparison, the empirical approximations of \cite{Dejager1988} predict 
$\log_{10} (\dot{M}/M_\odot/\hbox{year}) \simeq -5.8 \pm 0.5$ for
a red supergiant wind, roughly matching the rates needed if the 
wind is producing dust at the time of the explosion. 

The density of the wind is also tied to the expected phenomenology
of the explosion.  If we assume the standard $\rho_e \propto v^{-12}$ 
outer ejecta density profile for red supergiants (\citealt{Matzner1999})
then the expected shock velocity is 
\begin{equation}
   v_s = 8200 E_{51}^{9/20} M_{e10}^{-7/20} \dot{M}_{-4}^{-1/10} v_{w10}^{1/10}
        t_1^{-1/10}~\hbox{km/s}
\end{equation}
(e.g. \citealt{Chevalier2003}), where the total energy of the supernova is
$E= 10^{51} E_{51}$~ergs, the ejected mass is $M_e = 10 M_{e10}M_\odot$,
$\dot{M}=10^{-4} \dot{M}_{-4}M_\odot$/year and $t_1$ is the elapsed time
in days.  A shock expanding through a dense medium generates a luminosity
of $L_S = (1/2)\dot{M} v_s^3/v_w$, which we report in Table~\ref{tab:sedmodels}
for $v_s=5000$~km/s and $v_w=10$~km/s.  The emission from the forward shock
is usually too hard to be easily detected, but assuming that the reverse shock is 
cooling and that its softer emissions dominate the observable X-ray emissions, 
the expected X-ray luminosity is
\begin{equation}
   L_x \simeq { 9 \dot{M} v_s^3 \over 500 v_w} 
    \simeq 1.63 \times 10^7 E_{51}^{27/20} M_{e10}^{-21/20} \dot{M}_{-4}^{7/10}
                v_{w10}^{-7/10} t_1^{-3/10} L_\odot
    \label{eqn:xlum}
\end{equation}
with a temperature of order 1~keV (e.g. \citealt{Chevalier2003}). If we model
the X-ray fluxes in Table~\ref{tab:xray} using Eqn.~\ref{eqn:xlum} and a range
of additional absorption from $N_H=10^{20}$ to $10^{23}$~cm$^{-2}$, the mass 
loss rates for epochs 1, 3 and 5 (which have the smallest uncertainties) are
\begin{equation}
   \dot{M} \sim \left( 10^{-6.4\pm0.4}, 10^{-5.8\pm0.3},
    10^{-4.7\pm 0.3}\right) v_{w10} M_{e10}^{3/2} E_{51}^{-27/14} M_\odot/\hbox{year},
\end{equation}
respectively, with some evidence that the amount of excess absorption needed
above Galactic is increasing with time.   Such fluctuations and trends are
not unusual (e.g. \citealt{Dwarkadas2012}), but the X-ray emission appears
to be broadly consistent with the presence of a wind with roughly the right 
density to explain the extinction of the progenitor.  
\cite{Stockdale2012} and \cite{Yadav2012} report rising 
$\simeq 20$~GHz radio fluxes of $0.160\pm0.025$  and $0.315\pm0.018$~mJy roughly 
7 and 13 days after discovery that also argue for a significant wind at the
time of the SN.
If we model the radio emission following  \cite{Soderberg2005} assuming the
ejecta mass is $M_e=(15-1.4)M_\odot=13.6 M_\odot$ and $E_{51}=1$, we obtain
estimates of $\dot{M} \sim 10^{-5.0} v_{w10} M_\odot$/year, although given only two
data points at essentially the same frequency, the models are not tightly
constrained.  The Thompson optical depth of the wind is always negligible,
since it is a small ($\ltorder 1\%$) fraction of the dust optical depth,
but the cold dust solutions would likely convert the SN into a Type~IIn
because the H$\alpha$ luminosity from recombination 
is of order $3000R_{in15}L_\odot$ and increases linearly 
with the distance to the circumstellar material.  Thus, while the data
is fragmentary, the simplest interpretation appears to be that there was
a relatively steady $\dot{M} \sim 10^{-5.5}$ to $10^{-5.0}M_\odot$/year wind creating the
obscuration at the time of the SN.     

Since the primary reason for inappropriately using Galactic extinction
laws for circumstellar dust is almost certainly their ease of use, we supply in 
Table~\ref{tab:models} equivalently easy to use models for absorption
by circumstellar dust.   The problem is 
somewhat more complex because the extinction depends on both wavelength
and optical depth, but the absorption in the DUSTY models from the
UV to mid/near-IR ($0.3\mu$m to $5.0\mu$m) can be well modeled by
the functional form
\begin{equation}
   A_\lambda (\tau_V) = \tau_V \lambda^{-x} \sum_i \sum_j
          a_{ij} \tau_V^i \lambda^{-j}.
\end{equation}
for optical depths up to $\tau_V=20$.  Here $\lambda$ is the wavelength
in microns and $\tau_V$ is the total (absorption plus scattering) optical
depth in the V band.  Table~\ref{tab:models} provides these models for
$R_{out}/R_{in}=2$ and $10$ in a format where they can simply be extracted 
from the electronic paper using a mouse and inserted into most numerical environments.  
The fits reproduce the DUSTY results with rms fractional
residuals of 1.4\% (1.6\%) and 1.8\% (2.2\%) for the 
graphitic and silicate models and $R_{out}/R_{in}=10$ ($2$), although this includes 
some numerical rounding errors in the
DUSTY models at low optical depths and longer wavelengths.  They can
be extrapolated to longer wavelengths relatively safely, but at these
longer wavelengths one would almost always also need to include dust
emission.  {\it They should not be extrapolated to shorter wavelengths or
higher optical depths.} Fig.~\ref{fig:opdepth} shows the circumstellar
extinction of the DUSTY models used to build these interpolating functions
and the interpolating functions as extracted from Table~\ref{tab:models}
for $R_{out}/R_{in}=10$.  In some circumstances it will also be useful
to separate the direct and scattered light.  The fraction of the 
observed flux that is direct, $F_\lambda^{abs}$, can be well-modeled by 
\begin{equation}
   G_\lambda^{abs} = -2.5 \log_{10} F_\lambda^{abs} (\tau_V) = \tau_V^{1/2} (1+\lambda^{x})^{-1} \sum_i \sum_j
          a_{ij} \tau_V^{i/2} \lambda^{-j}.
     \label{eqn:direct}
\end{equation}
These interpolating functions are supplied in Table~\ref{tab:direct}
and the quality of the fits is shown in Fig.~\ref{fig:direct}.
For a source with intrinsic spectrum $S_\lambda$, the observed spectrum
is $S_\lambda 10^{-0.4 A_\lambda}$, the directly escaping flux is
$S_\lambda 10^{-0.4 A_\lambda} F_\lambda^{abs} = S_\lambda 10^{-0.4(A_\lambda+G_\lambda^{abs})}$ 
and the escaping but
scattered flux is $S_\lambda 10^{-0.4 A_\lambda} (1-F_\lambda^{abs})$.
Obviously, this approach can be generalized to other dust compositions
or size distributions.  We parametrize the models by $\tau_V$ because
$\tau_V$ is closely related to the physical properties of the wind (Eqn.~\ref{eqn:mdot}),
while relating it to a color (i.e. $E(B-V)$) would divorce the model from the
underlying physics.  There is no comparably simple means of 
treating dust emission because it depends critically on the input 
spectrum.

These differences are quantitative rather than qualitative -- in
circumstellar dust models the progenitor of SN~2012aw is still
fairly heavily obscured, as found by \cite{Fraser2012} and \cite{Vandyk2012},
and \cite{Walmswell2012} are correct that circumstellar dust introduces
 a bias that must be modeled when considering the statistics of 
SN progenitors, particularly when modeling non-detections in
analyses like \cite{Smartt2009b}.   However,
since changes in extinction physics exponentially modify quantities
like luminosities, these differences between the two dust geometries
are quantitatively important.  We note,
however, that SN where circumstellar dust will strongly bias inferences about the
progenitor have densities of circumstellar material that will lead
to X-ray or radio emission, as observed for SN~2012aw, because the
optical depth is proportional to the wind density.  For example,
combining Eqns.~\ref{eqn:mdot} and  \ref{eqn:xlum}, we see that
the expected X-ray luminosity is $L_X \propto \tau_V$.  This 
means that the properties of the explosion can be used to constrain
biases from dust around the progenitor even if the observations of
the progenitor are inadequate to constrain the circumstellar 
extinction.
   
There are, of course, additional complexities coming from the
geometry of the dust around the star that can lead to differences in the
effective optical depth along the line of sight for direct emission and
averaged over a significant fraction of the shell for the scattered 
emission.  Our simple models provide two means of approximating
some of these effects.  First, changes in the shell thickness
can be used to adjust the balance between scattered and absorbed light.
Second, the emergent flux can be modeled as a sum of direct light
with one optical depth, and scattered light with another in order to model
differences between the mean and line-of-sight optical depths.
{\it Independent of these questions, interstellar
and circumstellar extinction are quantitatively different, and in this
case ignoring the differences leads to a significant overestimate of 
the progenitor luminosity and mass. }

\acknowledgements

We thank J.~Beacom, J.~Eldridge, S.~Smartt, K.Z.~Stanek and T.A. Thompson 
for discussions and comments.  CSK is supported by NSF grant AST-0908816
and RK is supported by NSF grant AST-1108687.
This work is based in part on observations made with the Spitzer Space
Telescope, which is operated by the Jet Propulsion Laboratory, California
Institute of Technology under a contract with NASA.

{\it Facilities:} \facility{HST, SST}

\vfill\eject 

\begin{deluxetable}{ccrrrrrl}
\tabletypesize{\scriptsize}
\tablecaption{Swift X-ray Data\label{tab:xray}}
\tablehead{
Obsid &$\langle\hbox{Date}\rangle$ &$T_{exp}$ (sec) &0.2--10~keV &0.2--0.5~keV &0.5--2~keV &2--10~keV &Range
 }
\startdata
         00032319001 & March  21.0  &$  9151 $ &$  1.781 \pm  0.536 $ &$  0.174 \pm  0.197  $ &$  1.445 \pm  0.458 $ &$  0.162 \pm  0.197 $   &03/21 \\ 
         00032319002 & March  23.0  &$  5856 $ &$  0.449 \pm  0.376 $ &$ -0.090 \pm  0.028  $ &$  0.369 \pm  0.306 $ &$  0.171 \pm  0.217 $   &03/23  \\
         00032319003 & March  25.0  &$  8843 $ &$  1.528 \pm  0.497 $ &$ -0.053 \pm  0.018  $ &$  1.520 \pm  0.475 $ &$  0.061 \pm  0.145 $   &03/25  \\
    00032319004--006 & March  30.4  &$ 22714 $ &$  0.618 \pm  0.229 $ &$  0.037 \pm  0.112  $ &$  0.323 \pm  0.146 $ &$  0.261 \pm  0.136 $   &03/28-04/01 \\ 
    00032315015--030 & April  12.5  &$ 48188 $ &$  0.358 \pm  0.136 $ &$  0.117 \pm  0.089  $ &$  0.059 \pm  0.058 $ &$  0.184 \pm  0.085 $   &04/01-04/22  \\
    00032315031--042 & June   14.9  &$ 17430 $ &$ -0.123 \pm  0.198 $ &$ -0.302 \pm  0.080  $ &$  0.280 \pm  0.164 $ &$ -0.100 \pm  0.076 $   &04/26-07/14  \\
\enddata
\tablecomments{$\langle\hbox{Date}\rangle$ is the exposure time weighted average date, where Range gives the
  range of dates spanned by the epochs.  The counts are in units of counts/ksec. }
\end{deluxetable}

\def\hp2{\hphantom{0}}
\begin{deluxetable}{lrcccccc}
\tabletypesize{\scriptsize}
\tablecaption{Model Fits \label{tab:sedmodels}}
\tablehead{
Type &$T_d$ (K) &$\log_{10}(L_*/L_\odot)$ &$T_*$ (K) &$\log_{10}\tau_V$  &$\log_{10} (R_{in}/\hbox{cm})$ &$\log_{10}\dot{M}$ &$\log_{10}(L_s/L_\odot)$
 }
\startdata
 sil &$  100$
&$    4.99 \hp2 (    4.88/    5.12)$
&$3754 \hp2 (3483/4168)$
&$    0.87 \hp2 (    0.72/    0.97)$
&$   16.97 \hp2 (   16.84/   17.10)$
&$   -2.85 \hp2 (   -3.14/   -2.63)$
&$    9.16 \hp2 (    8.87/    9.38)$
\\ 
 sil &$  500$
&$    4.93 \hp2 (    4.80/    5.06)$
&$3600 \hp2 (3357/3991)$
&$    0.80 \hp2 (    0.48/    0.93)$
&$   15.31 \hp2 (   15.13/   15.46)$
&$   -4.59 \hp2 (   -5.09/   -4.31)$
&$    7.43 \hp2 (    6.92/    7.70)$
\\ 
 sil &$ 1000$
&$    4.91 \hp2 (    4.81/    5.01)$
&$3603 \hp2 (3385/3976)$
&$    0.79 \hp2 (    0.55/    0.91)$
&$   14.80 \hp2 (   14.67/   14.93)$
&$   -5.11 \hp2 (   -5.48/   -4.86)$
&$    6.90 \hp2 (    6.53/    7.15)$
\\ 
 sil &$ 1500$
&$    4.90 \hp2 (    4.81/    4.97)$
&$3823 \hp2 (3473/4264)$
&$    0.85 \hp2 (    0.67/    0.94)$
&$   14.50 \hp2 (   14.36/   14.60)$
&$   -5.35 \hp2 (   -5.67/   -5.17)$
&$    6.66 \hp2 (    6.35/    6.85)$
\\ 
 gra &$  100$
&$    5.14 \hp2 (    4.96/    5.30)$
&$3897 \hp2 (3528/4308)$
&$    0.64 \hp2 (    0.45/    0.74)$
&$   17.40 \hp2 (   17.25/   17.53)$
&$   -2.65 \hp2 (   -3.00/   -2.43)$
&$    9.36 \hp2 (    9.01/    9.58)$
\\ 
 gra &$  500$
&$    4.95 \hp2 (    4.82/    5.08)$
&$3528 \hp2 (3360/3821)$
&$    0.44 \hp2 (    0.13/    0.60)$
&$   15.81 \hp2 (   15.70/   15.93)$
&$   -4.46 \hp2 (   -4.87/   -4.18)$
&$    7.56 \hp2 (    7.14/    7.84)$
\\ 
 gra &$ 1000$
&$    4.89 \hp2 (    4.79/    4.97)$
&$3562 \hp2 (3372/3910)$
&$    0.43 \hp2 (    0.17/    0.57)$
&$   14.97 \hp2 (   14.88/   15.07)$
&$   -5.30 \hp2 (   -5.65/   -5.07)$
&$    6.71 \hp2 (    6.36/    6.95)$
\\ 
 gra &$ 1500$
&$    4.84 \hp2 (    4.76/    4.93)$
&$3694 \hp2 (3392/4222)$
&$    0.48 \hp2 (    0.16/    0.61)$
&$   14.49 \hp2 (   14.37/   14.59)$
&$   -5.74 \hp2 (   -6.17/   -5.50)$
&$    6.28 \hp2 (    5.84/    6.51)$
\\ 
\enddata
\tablecomments{These are the median and 90\% confidence intervals of the MCMC results. The implied
  mass loss rate $\dot{M}$ is in units of $v_{w10}\kappa_{V100}^{-1}M_\odot/\hbox{year}$,
  and the shock luminosity $L_s$ is in units of 
  $(v_s/5000\hbox{km/s})^3\kappa_{V100}^{-1} L_\odot$.
  }
\end{deluxetable}

\def\hp{\hphantom{000}}
\begin{deluxetable}{l}
\tablecaption{Graphitic and Silicate Circumstellar Extinction Laws}
\tablehead{\multicolumn{1}{c}{Models with $R_{out}/R_{in}=2$}}
\tabletypesize{\small}
\tablewidth{0pt}
\startdata
\hp $tgra1=(   0.500446+   1.795729*t   -1.877658*t*t+   0.852820*t**3   -0.141635*t**4)$\\
\hp $tgra2=(   4.318269  -14.236698*t+  13.804110*t*t   -5.991369*t**3+   0.959539*t**4)/l$\\
\hp $tgra3=(  -5.114167+  32.462564*t  -26.895305*t*t+  10.197398*t**3   -1.414338*t**4)/l**2$\\
\hp $tgra4=(   3.384105  -21.107633*t+  12.229167*t*t   -2.172318*t**3   -0.149866*t**4)/l**3$\\
\hp $tgra5=(  -1.059677+   5.553703*t   -1.527415*t*t   -0.881450*t**3+   0.391763*t**4)/l**4$\\
\hp $tgra6=(   0.121772   -0.518968*t   -0.048566*t*t+   0.248002*t**3   -0.077054*t**4)/l**5$\\
 $agraphite2 = (tgra1+tgra2+tgra3+tgra4+tgra5+tgra6)*t*l**(    -1.375229)$\\
\hline
\hp $tsil1=(   0.437549   -0.446323*t+   0.648423*t*t   -0.321970*t**3+   0.055555*t**4)$\\
\hp $tsil2=(  -0.486741+   4.034854*t   -5.530127*t*t+   2.711095*t**3   -0.469112*t**4)/l$\\
\hp $tsil3=(   1.166512  -12.015845*t+  14.917191*t*t   -7.058630*t**3+   1.204954*t**4)/l**2$\\
\hp $tsil4=(  -0.655682+  13.849514*t  -14.342367*t*t+   6.165157*t**3   -0.984406*t**4)/l**3$\\
\hp $tsil5=(   0.169689   -4.956815*t+   4.137525*t*t   -1.419899*t**3+   0.176166*t**4)/l**4$\\
\hp $tsil6=(  -0.016829+   0.582619*t   -0.381166*t*t+   0.083593*t**3   -0.002153*t**4)/l**5$\\
 $asilicate2 = (tsil1+tsil2+tsil3+tsil4+tsil5+tsil6)*t*l**(    -0.642318)$\\
\hline
\multicolumn{1}{c}{Models with $R_{out}/R_{in}=10$} \\
\hline
\hp $tgra1=(   0.760499+   0.879164*t   -0.350748*t*t   -0.039612*t**3+   0.034161*t**4)$\\
\hp $tgra2=(   4.061343   -7.166933*t+   2.791544*t*t+   0.214647*t**3   -0.233685*t**4)/l$\\
\hp $tgra3=(  -5.133851+  16.344656*t   -4.283100*t*t   -1.764900*t**3+   0.780217*t**4)/l**2$\\
\hp $tgra4=(   3.387184  -10.066016*t   -1.260999*t*t+   4.103272*t**3   -1.160204*t**4)/l**3$\\
\hp $tgra5=(  -1.052057+   2.479576*t+   1.618868*t*t   -2.030708*t**3+   0.516503*t**4)/l**4$\\
\hp $tgra6=(   0.120327   -0.214118*t   -0.293914*t*t+   0.295687*t**3   -0.071840*t**4)/l**5$\\
 $agraphite10 = (tgra1+tgra2+tgra3+tgra4+tgra5+tgra6)*t*l**(    -1.475236)$\\
\hline
\hp $tsil1=(   0.197398   -0.293417*t+   0.192686*t*t   -0.041375*t**3+   0.000902*t**4)$\\
\hp $tsil2=(   0.093593+   2.491030*t   -1.453387*t*t+   0.239280*t**3+   0.013273*t**4)/l$\\
\hp $tsil3=(   0.357331   -6.883382*t+   3.239407*t*t   -0.198182*t**3   -0.121949*t**4)/l**2$\\
\hp $tsil4=(   0.022567+   7.169214*t   -1.884278*t*t   -0.728088*t**3+   0.309664*t**4)/l**3$\\
\hp $tsil5=(  -0.065599   -2.173130*t   -0.305525*t*t+   0.839291*t**3   -0.223389*t**4)/l**4$\\
\hp $tsil6=(   0.012188+   0.216324*t+   0.133118*t*t   -0.153598*t**3+   0.036469*t**4)/l**5$\\
 $asilicate10 = (tsil1+tsil2+tsil3+tsil4+tsil5+tsil6)*t*l**(    -0.323043)$\\
\enddata
\tablecomments{These expressions are designed to simply be grabbed with
 a mouse from the electronic paper and inserted into most numerical 
 environments.  The input quantities are $t=\tau_V/10$ and $l=\lambda$
 in microns, and the output quantity is $A_\lambda(\tau_V)$.
 They are valid for $\lambda \geq 0.3\mu$m
 and $\tau_V \leq 20$ and should not be extrapolated outside this
 range.  No emission by the dust is included in the models.
   }
\label{tab:models}
\end{deluxetable}

\begin{deluxetable}{l}
\tablecaption{Direct Emission Fractions for Graphitic and Silicate Circumstellar Extinction Laws}
\tablehead{\multicolumn{1}{c}{Models with $R_{out}/R_{in}=2$}}
\tabletypesize{\small}
\tablewidth{0pt}
\startdata
\hp $ggra1=(   0.091091-2.031641*s+   5.049221*s*s-6.399577*s**3+   2.086527*s**4)$\\
\hp $ggra2=(  -0.705931+   3.566326*s-40.821335*s*s+  48.029644*s**3-14.815801*s**4)/l$\\
\hp $ggra3=(   1.604726-22.761479*s+  90.223336*s*s-88.179577*s**3+  25.041643*s**4)/l**2$\\
\hp $ggra4=(  -1.240894+  20.383229*s-63.820599*s*s+  57.434801*s**3-15.475733*s**4)/l**3$\\
\hp $ggra5=(   0.392600-6.856117*s+  19.106438*s*s-16.340312*s**3+   4.240107*s**4)/l**4$\\
\hp $ggra6=(  -0.044443+   0.800158*s-2.078890*s*s+   1.716855*s**3-0.433098*s**4)/l**5$\\
 $ggraphite2 =-(ggra1+ggra2+ggra3+ggra4+ggra5+ggra6)*s/(1+l**(     3.608885))$\\
\hp $gsil1=(   0.016559-9.435716*s+   1.279347*s*s-0.617515*s**3+   0.351253*s**4)$\\
\hp $gsil2=(   0.054245+  15.096287*s+   1.441236*s*s-0.704551*s**3-0.049853*s**4)/l$\\
\hp $gsil3=(  -0.308782-6.642689*s-15.320068*s*s+  12.697774*s**3-3.365767*s**4)/l**2$\\
\hp $gsil4=(   0.367973-6.190985*s+  18.372915*s*s-12.435162*s**3+   2.599908*s**4)/l**3$\\
\hp $gsil5=(  -0.134091+   3.475205*s-5.810992*s*s+   2.902163*s**3-0.261672*s**4)/l**4$\\
\hp $gsil6=(   0.015684-0.468930*s+   0.577899*s*s-0.158431*s**3-0.045234*s**4)/l**5$\\
 $gsilicate2 =-(gsil1+gsil2+gsil3+gsil4+gsil5+gsil6)*s/(1+l**(     4.733179))$\\
\hline
\multicolumn{1}{c}{Models with $R_{out}/R_{in}=10$} \\
\hline
\hp $ggra1=(  -0.132213-7.859540*s-1.722856*s*s+   0.017627*s**3+   0.200712*s**4)$\\
\hp $ggra2=(   0.397806+   6.851899*s+   0.074252*s*s+  10.503128*s**3-3.322806*s**4)/l$\\
\hp $ggra3=(  -0.420577-7.736732*s+   5.079083*s*s-11.886283*s**3+   2.016602*s**4)/l**2$\\
\hp $ggra4=(   0.281626+   3.202032*s+   0.592135*s*s+   1.043499*s**3+   1.258519*s**4)/l**3$\\
\hp $ggra5=(  -0.093566-0.543338*s-1.283187*s*s+   1.231904*s**3-0.913404*s**4)/l**4$\\
\hp $ggra6=(   0.011334+   0.026020*s+   0.236427*s*s-0.256076*s**3+   0.140387*s**4)/l**5$\\
 $ggraphite10 =-(ggra1+ggra2+ggra3+ggra4+ggra5+ggra6)*s/(1+l**(     4.399311))$\\
\hp $gsil1=(   0.041183-10.239763*s+   3.360815*s*s-3.994570*s**3+   1.668688*s**4)$\\
\hp $gsil2=(  -0.127688+  19.263277*s-13.077187*s*s+  19.483457*s**3-8.060155*s**4)/l$\\
\hp $gsil3=(   0.141609-15.254131*s+  18.619615*s*s-32.651846*s**3+  14.198402*s**4)/l**2$\\
\hp $gsil4=(  -0.089791+   1.766552*s-13.867558*s*s+  28.015047*s**3-12.378082*s**4)/l**3$\\
\hp $gsil5=(   0.042445+   0.565792*s+   5.721465*s*s-10.944883*s**3+   4.706386*s**4)/l**4$\\
\hp $gsil6=(  -0.006950-0.107301*s-0.821527*s*s+   1.470719*s**3-0.616414*s**4)/l**5$\\
 $gsilicate10 =-(gsil1+gsil2+gsil3+gsil4+gsil5+gsil6)*s/(1+l**(     4.756651))$\\
\enddata
\tablecomments{These expressions are designed to simply be grabbed with
 a mouse from the electronic paper and inserted into most numerical
 environments.  The input quantities are $s=(\tau_V/10)^{1/2}$ (NOTE THE
 CHANGE FROM $t=\tau_V/10$ IN TABLE~\ref{tab:models}!) and $l=\lambda$
 in microns, and the output quantity is $G_\lambda^{abs}(\tau_V)$ from Eqn.~\ref{eqn:direct}.
 They are valid for $\lambda \geq 0.3\mu$m
 and $\tau_V \leq 20$ and should not be extrapolated outside this
 range.  No emission by the dust is included in the models.
   }
\label{tab:direct}
\end{deluxetable}

\end{document}